\documentclass[%
  prx,                   
  twocolumn,             
  superscriptaddress,    
  longbibliography,      
  nofootinbib,           
]{revtex4-2}

\usepackage[utf8]{inputenc}
\usepackage[T1]{fontenc}
\usepackage{lmodern}            
\usepackage{amsmath,amssymb}
\usepackage{graphicx}
\usepackage{subcaption}
\usepackage{hyperref}
\usepackage{xcolor}             
\usepackage{siunitx}            
\usepackage{mwe}
\usepackage{svg} 
\usepackage{dcolumn}
\usepackage{bm}
\usepackage{bbm}
\usepackage{amsmath}
\usepackage{amssymb}

\usepackage{comment}
\usepackage{adjustbox}
\usepackage{pifont}
\usepackage{ragged2e} 
\newcommand{\revDS}[1]{\color{black}{#1}}

\definecolor{darkgreen}{RGB}{0,100,0}
\newcommand{\alex}[1]{\color{black}{#1}}

\makeatletter
\long\def\@makecaption#1#2{%
  \vskip\abovecaptionskip
  \justifying\normalfont\small
  #1: #2\par
  \vskip\belowcaptionskip
}
\makeatother
\begin{document}

\title{Nonlinear competition avoidance favors coexistence in microbial populations}

\author{Mattia Mattei}
\affiliation{Departament d'Enginyeria Inform{\`a}tica i Matem{\`a}tiques, Universitat Rovira i Virgili, 43007 Tarragona, Spain}
\author{David Soriano-Paños}
\affiliation{Departament d'Enginyeria Inform{\`a}tica i Matem{\`a}tiques, Universitat Rovira i Virgili, 43007 Tarragona, Spain}
\affiliation{GOTHAM lab, Institute for Biocomputation and Physics of Complex Systems,\\ University of Zaragoza, 50018 Zaragoza, Spain}
\author{Alex Arenas}
\affiliation{Departament d'Enginyeria Inform{\`a}tica i Matem{\`a}tiques, Universitat Rovira i Virgili, 43007 Tarragona, Spain}
\affiliation
{ComSCIAM, Universitat Rovira i Virgili, 43007 Tarragona, Spain}

\begin{abstract}

Bacteria regulate their motility through a variety of mechanisms, including quorum sensing (QS) {\alex and other density--dependent responses mediated by diffusible signals.}
{\alex While nonlinear density-dependent motility is well known in active-matter theory to generate nonequilibrium spatial patterns, its consequences for the coexistence of growing, interacting species remain less explored.} 
Here we develop a minimal {\alex spatially structured} model for two strongly competing species {\alex in which local demographic interactions are coupled to an escape response: each species increases its motility nonlinearly {\alex (sigmoidal)} with the local abundance of its competitor.}
We show that {\alex this sigmoidal motility regulation}
promotes optimal spatial self-organization {\alex and can sustain long term coexistence via segregation, even in parameter regimes that yield competitive exclusion in well-mixed Lotka--Volterra dynamics.}
On two-dimensional lattices, the interplay between demographic competition and density--dependent motility
{\alex generates a range of emergent patterns, }
including regimes in which the weaker competitor counterintuitively {\alex has higher total abundance.} 
Overall, our results identify nonlinear, {\alex competitor--induced} motility as a fundamental mechanism capable of sustaining coexistence in competing microbial populations.

\end{abstract}

\maketitle

\section{Introduction}

Many bacterial species exhibit various forms of motility, ranging from swimming in liquid environments to coordinated swarming across solid substrates \cite{Wadhwa}. These behaviors are often regulated responses to environmental and physiological signals, mediated by mechanisms such as chemotaxis \cite{Budrene} or quorum sensing (QS) \cite{Waters}, whereby {\revDS the accumulated autoinducer signals reflect the local bacteria density and trigger coordinated changes in gene expression}. In several species, {\alex QS represses} motion by downregulating the flagellar apparatus or inducing adhesive phenotypes that favor surface attachment and reduced mobility \cite{Daniels, Hoang}. {\alex In contrast}, signaling can also stimulate motion: QS-controlled synthesis of biosurfactants such as surfactin in \emph{Bacillus subtilis} \cite{Kinsinger}, putisolvins in \emph{Pseudomonas putida} \cite{Carcamo}, or AHL-mediated responses in \emph{Rhizobium etli} \cite{Daniels2} has been shown to promote swarming, surface expansion, and {\alex microcolonization dispersal}. All these mechanisms correspond to changes in motility driven by the local concentration of signaling molecules, which reflect the density of the nearby cells that synthesize them. 

From a theoretical perspective,{ \alex density-dependent motility has been a central theme in active-matter physics because it can generate spatial structure in the absence of explicit attractive interactions. Tailleur and Cates \cite{Tailleur} showed that run-and-tumble particles which reduce their speed in high density regions can spontaneously separate into dense and dilute phases, a phenomenon now referred to as motility-induced phase separation (MIPS) \cite{Cates}. Subsequent work has extended these ideas to multi-species systems and to more general forms of density-dependent motility, revealing a rich spectrum of nonequilibrium patterns, including clusters, concentric rings, and propagating bands} \cite{Curatolo, Dinelli1, Duan, Ridgway, Dinelli2, Mattei2}.  


Although {\alex density-dependent motility is well established as a pattern formation mechanism, its consequences for survival and coexistence in growing multispecies communities remain comparatively less explored }
\cite{Hallatschek_review}. This gap is notable given {\alex the challenge of explaining the high diversity observed in microbial communities} 
\cite{Coyte}. Coexistence among species competing for the same resources typically requires some form of niche differentiation \cite{Chesson}, and such differences may arise from an interplay between local competitive dynamics and spatial processes operating at larger scales. Even in spatially homogeneous environments, coexistence can be achieved through competition–colonization trade-offs, in which superior competitors are poorer colonizers of empty space than inferior ones \cite{Amarasekare}. Spatial segregation maintained by escape or fugitive strategies \cite{Bolker} of the weaker competitor has been repeatedly documented in ecology \cite{Fuller, Webster, Durant}.

This raises the question {\alex of whether density--dependent motility can itself generate an effective competition--colonization tradeoff in microbial systems}.
{\alex Previous studies have combined density-dependent diffusivity with population dynamics, largely emphasizing the resulting spatial patterns.}
Cates et al. \cite{Cates2} investigated this in a single species undergoing logistic growth, showing that phase separation can still arise, although growth eventually arrests the coarsening. Curatolo et al. \cite{Curatolo} analyzed density-dependent motility in a pair of competing bacterial strains, reporting demixing and co-localization patterns, yet within a framework where the species are indistinguishable.

\begin{figure*}[t!]
    \centering
    \includegraphics[width=2\columnwidth]{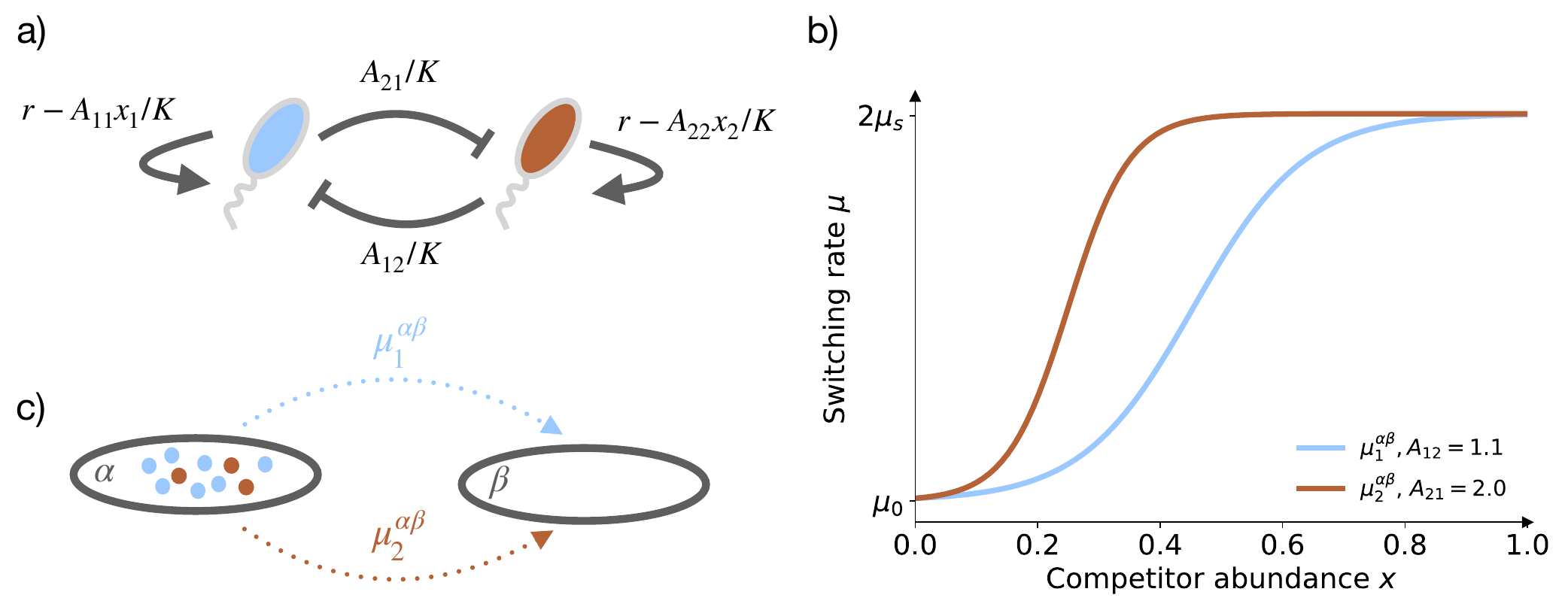}
    \caption{Schematic illustration of the model.  (a) Demographic dynamics within each patch, including intrinsic growth at rate $r$, carrying capacity $K$, and intra- and inter-specific competition terms $A_{ii}$ and $A_{ij}$, respectively. (b) Sigmoidal escape response functions [eq.~\ref{eq:mu}] for the species 1 with $A_{12}=1.1$ (blue line) and for species 2 for $A_{21}=2.0$ (red line), considering $\Delta=0.5$ and $\xi=10$. (c) Initial configuration of the system in the case of the two-patches setting, in which both species are initially present in patch $\alpha$ and may subsequently migrate to patch $\beta$ with species-specific migration rates $\mu_1^{\alpha\beta}$ and $\mu_2^{\alpha\beta}$.}
    \label{fig1}
\end{figure*}

{\alex Here we focus instead on coexistence between asymmetrically competing species that would undergo competitive exclusion in a well-mixed Lotka--Volterra setting. }
We develop a minimal description for two strongly competing species in which {\alex local demographic interactions are coupled to an escape response: each species increases its motility nonlinearly with the local abundance of its competitor, with the strength of the response tied to the strength of interspecific competition.}

{\alex We show that this coupling can promote  coexistence across broad parameter regimes.}
{\alex We first characterize a two-patch setting analytically, isolating how the functional form of the motility response controls the emergence of coexistence. Then, we extend the analysis to two-dimensional lattices where the same mechanism generates diverse spatial patterns, including regimes in which the weaker competitor occupies a larger spatial region than its stronger rival.} 


\section{Results}

\subsection{The Model}

The model is a standard extension of the competitive Lotka–Volterra equations for two species, labeled by indices $1$ and $2$, to a spatially structured system. The full system consists of $2m$ coupled equations, where $m$ is the total number of spatial units, or patches. For species 1 in patch $\alpha$, the dynamics reads
\begin{equation}
\begin{split}
\frac{d x_1^{\alpha}(t)}{dt} &=
x_1^{\alpha}(t)\!\left[
r
- \frac{A_{11}}{K}\,x_1^{\alpha}(t)
- \frac{A_{12}}{K}\,x_2^{\alpha}(t)
\right] \\
&\quad
+ \sum_{\beta}
\Big[
\mu_1^{\beta\alpha}(t)\,x_1^{\beta}(t)
-
\mu_1^{\alpha\beta}(t)\,x_1^{\alpha}(t)
\Big].
\end{split}
\label{eq:model}
\end{equation}
The first line represents the demographic dynamics: $r$ is the intrinsic growth rate, $K$ the carrying capacity, and $A_{11}$ and $A_{12}$ quantify intraspecific limitation and interspecific competition, respectively. This is schematically represented in Figure \ref{fig1}a. 
{\alex Here $x_i^{\alpha}(t)$ denotes the density of species $i$ in patch $\alpha$, so that $r$ has units of inverse time, $K$ is a carrying density with the same units as $x$, and $\mu_i^{\alpha\beta}(t)$ are per-capita transfer rates with units of inverse time. With the rescalings $x \to x/K$ and $t \to r t$, the coefficients $A_{ij}$ are dimensionless; throughout we work in these rescaled units and set $r = K = A_{11} = A_{22} = 1$, focusing on the roles of interspecific competition and motility.
}

\begin{figure*}[t!]
   \centering
    \includegraphics[width=2.0\columnwidth]{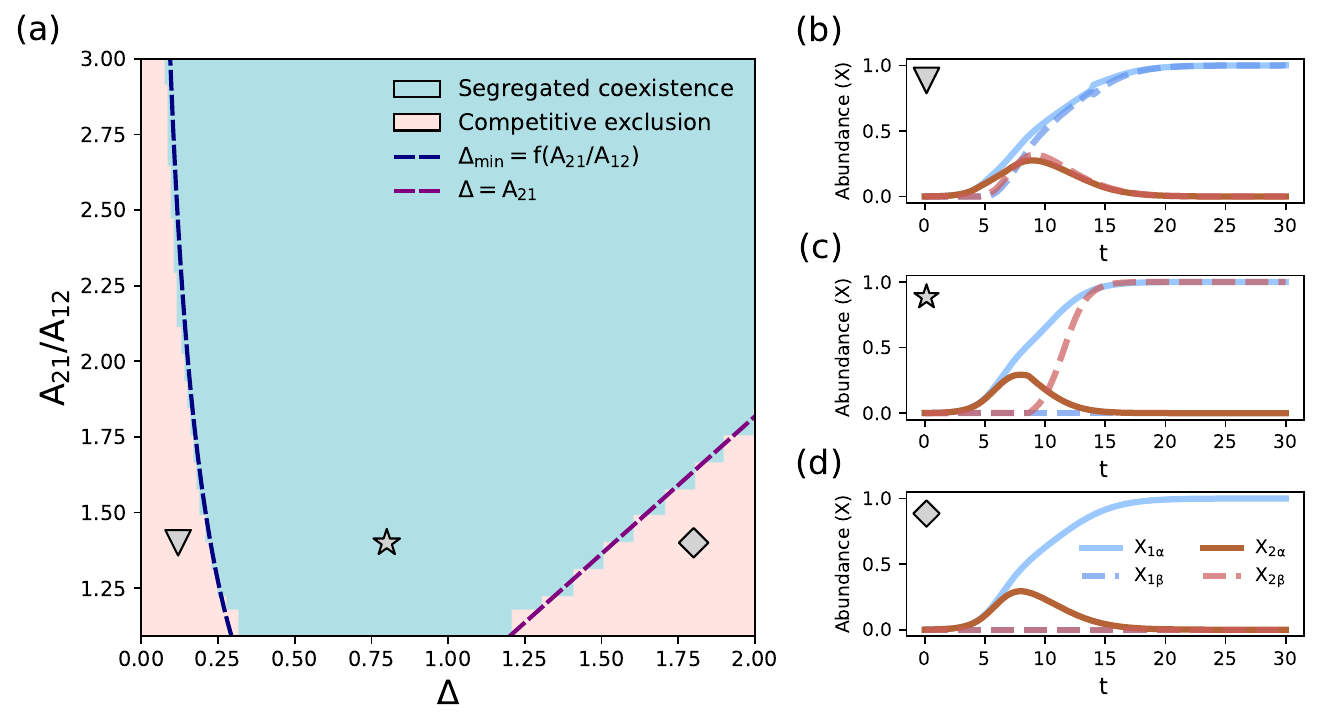}
    \caption{(a) Phase diagram distinguishing segregated coexistence from competitive exclusion as a function of the motility threshold $\Delta$ and the interaction ratio $A_{21}/A_{12}$ for two neighboring patches, in the theoretical limits $\xi \to \infty$ and $\mu_0 \to 0$. The background heatmap is obtained from numerical integration of Eqs.~\ref{eq:model}, while dashed lines indicate analytical predictions (see Appendix~A). The three markers correspond to the parameter values $A_{21}/A_{12}=1.4$ and $\Delta = 0.12,\ 0.8,\ 1.8$, for which representative temporal dynamics are shown in panels (b–d).  (b–d) Representative time series illustrating the different dynamical outcomes. Symbol \ding{116} denotes competitive exclusion resulting from the migration of both species to patch $\beta$; \ding{72} indicates coexistence driven by the migration of species~2; and \ding{117} corresponds to competitive exclusion when neither species migrates. All simulations use $A_{11}=A_{22}=K=r=1$, $\mu_S=0.1$, and $A_{12}=1.1$.
}
    \label{fig2}
\end{figure*}

{\revDS The second line encodes motility between neighboring patches, being $\mu_1^{\alpha\beta}(t)$ the rate at which species 1 move from $\alpha$ to $\beta$ at time $t$. This rate reads:
\begin{equation}
\begin{split}
\mu_1^{\alpha\beta}(t)
= M_{\alpha\beta}\Bigg[\mu_0
+ 
\frac{2\,\mu_S}{1 + \exp{\big[{\xi(\Delta - A_{12}\,x_2^{\alpha}}(t))}\big]}\Bigg],
\label{eq:mu}
\end{split}
\end{equation}}
{\revDS where we have expressed species motility as the combination of a constant diffusivity $\mu_0$ and a nonlinear sigmoidal modulation that increases with the local abundance of the competitor}. The choice of a sigmoid function is common when modelling processes such as quorum sensing \cite{Curatolo, Dinelli2, Mattei, Mattei2} where reaching a critical density threshold triggers a qualitative change. {\revDS In our case, such response is an ``escape'' response, moving a species apart from the areas with high local density of its competitor}.
The parameter $\xi$ controls the steepness of this response, while $\Delta$ sets its inflection point, that can be seen as an activation threshold for $\xi\to\infty$. Note that the abundance of competitors enters the motility function multiplied by the interaction coefficient $A_{12}$. This choice guarantees that the influence of one species on the {\revDS growth of the} other is consistently mirrored in the modulation of its motility, thereby coupling demographic and escape pressures. This functional form of the escape response is plotted in Figure~\ref{fig1}b, where we show how the competition coefficients $A_{21}$ and $A_{12}$ affect the steepness and the inflection point of the sigmoid for both species. Finally, $\mathbf{M}$ is an $m \times m$ symmetric matrix that encodes the spatial adjacency between patches. In the two-patch scenario, we simply have $M_{\alpha\beta} = M_{\beta\alpha} = 1$. For lattice geometries, neighboring sites satisfy $M_{\alpha\beta} = M_{\beta\alpha} = 1/\mathcal{N}$, where $\mathcal{N}$ denotes the number of nearest neighbors, while all other entries are zero.
{\alex
Throughout, we understand coexistence as long-term persistence of both species at the metapopulation level, i.e., $x_1=\sum_\alpha x_1^\alpha>0$ and $x_2=\sum_\alpha x_2^\alpha>0$ as $t\to\infty$. In the competitive-exclusion regime $A_{12}>1$ and $A_{21}>1$, stable coexistence within a single patch is impossible in the classical Lotka--Volterra dynamics, so coexistence in our spatial system necessarily relies on spatial segregation and/or sustained spatial patterning.
}
In what follows, we begin with the simplified setting consisting of two neighboring patches, and later we extend the analysis to two-dimensional lattices, which can be seen as discrete representations of an underlying continuous space.

\subsection{Two Patches}

{\revDS
Let us consider two connected patches, $\alpha$ and $\beta$, and assume an initial condition where a small seed of both species is present only in patch $\alpha$, while patch $\beta$ is empty (Figure~\ref{fig1}c). 
Regarding the demographic interactions of the species, we assume the regime of competitive exclusion, where, in our setup, the interaction rates must fulfill $A_{12}>1$, $A_{21}>1$ and $A_{12}\neq A_{21}$. Within the classical Lotka–Volterra framework, stable coexistence of both species within a single patch is impossible in this regime. Our goal is therefore to investigate whether the nonlinear motility response introduced in Eq.~\ref{eq:mu} can generate the conditions required for coexistence across the two patches. To address this question, let us first explore the scenario in the limits $\mu_0 \to 0$ and $\xi \to \infty$, thus neglecting the constant bacterial diffusion and considering a step function triggering species motility when the density of the competitor exceeds $\Delta$. 
We will study later the effects of relaxing both assumptions.

In Fig.~\ref{fig2}a we show the phase diagram of the system as a function of the ratio between the interaction rates $A_{21}/A_{12}$ and the motility threshold $\Delta$. The resulting phase diagram reveals a broad coexistence region, which arises specifically from the structure of our motility function. As both species cannot coexist within a single patch, the only viable configuration for both species to thrive is spatial segregation: one species ultimately colonizes one patch while the other species colonizes the other. Under the initial conditions depicted in Fig.~\ref{fig1}c, {\alex species~2 must cross its motility activation condition, escape patch $\alpha$ and colonize patch $\beta$, while species~1 must never activate its escape response, preventing it from invading both patches.} Mathematically, if $A_{21}>A_{12}$ this occurs when 
\begin{equation}
x_2^\alpha(t) < \frac{\Delta}{A_{12}} \qquad \forall t\in[0,\infty],
\label{eq:cond1}
\end{equation}
so species $1$ always stays in $\alpha$ and there exists a time $\tau$ for which, 
\begin{equation}
x_1^\alpha(\tau) \ge \frac{\Delta}{A_{21}},
\label{eq:cond2}
\end{equation}
thus triggering the motility of species $2$.

For a fixed value of the competition ratio $A_{21}/A_{12}$, increasing the motility threshold $\Delta$ leads to two successive transitions: first from competitive exclusion to coexistence, and then from coexistence back to exclusion. The first transition separates a regime in which the dominant species remains mobile and ultimately colonizes both patches, so that condition~(\ref{eq:cond1}) is violated, from a segregated coexistence state in which each species occupies a different patch and both conditions are satisfied. In Appendix~A we analytically derive the boundary between these two regimes, shown as the blue curve in Fig.~\ref{fig2}c. This curve defines the minimum activation threshold, $\Delta_{\min}=f(A_{21}/A_{12})$, required for condition~(\ref{eq:cond1}) to hold. In Figs.~\ref{fig2}b-c we show the time evolution of the abundances of both species in both patches in these first two regimes. 

Further increasing $\Delta$ prevents the weak species from escaping from patch $\alpha$, as Eq.~(\ref{eq:cond2}) is no longer satisfied. Consequently, none of the species invades patch $\beta$ and the weak species becomes extinct in patch $\alpha$, as shown in Fig.~\ref{fig2}d. The boundary can be found analytically by setting $x_1=K$ in Eq.~\ref{eq:cond2}, yielding $\Delta=A_{21}$ as $K=1$.
\begin{figure}[t!]
    \centering
    \hspace*{-4mm}
    \includegraphics[width=1.0\columnwidth]{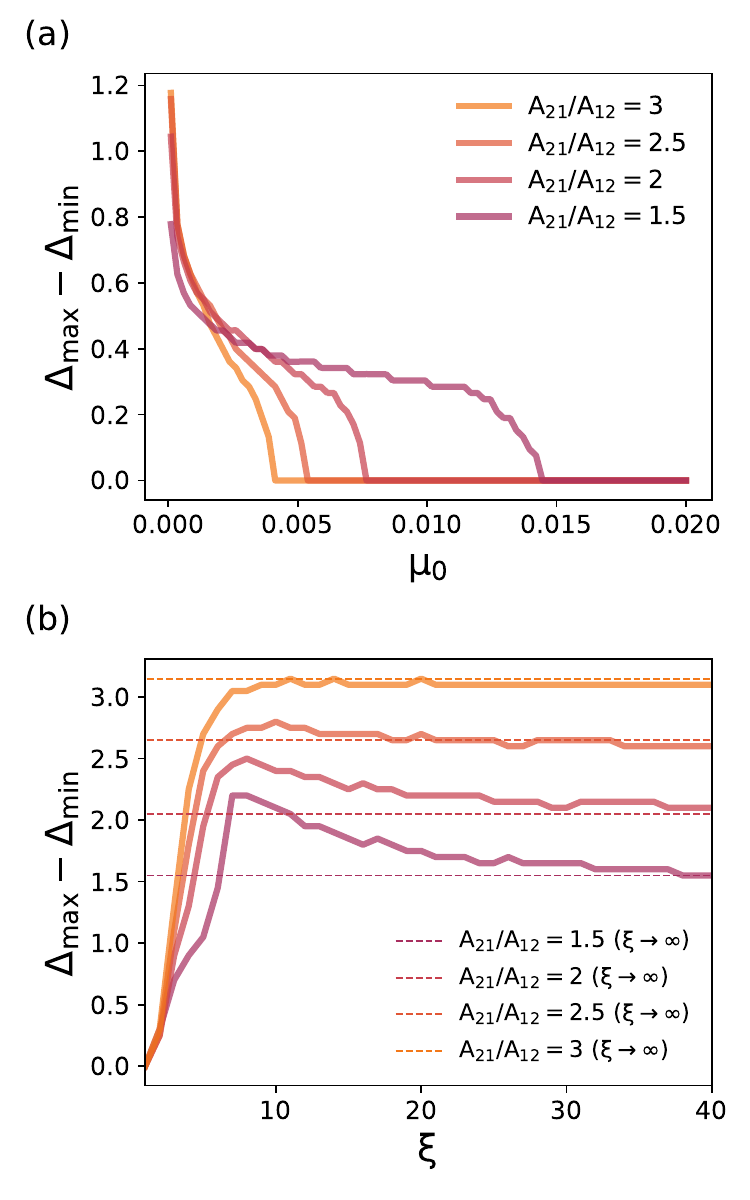}
    \caption{(a) Range of motility thresholds, $\Delta_{\text{max}} - \Delta_{\text{min}}$, that permit segregated coexistence as a function of the linear diffusion rate $\mu_0$, shown for different values of the interspecific competition ratio $A_{21}/A_{12}$. (b) Range of motility thresholds, $\Delta_{\text{max}} - \Delta_{\text{min}}$, as a function of the steepness parameter $\xi$ of the sigmoidal escape response, again for fixed values of $A_{21}/A_{12}$. The dashed lines indicate the corresponding asymptotic values obtained in the limit $\xi \to \infty$. All curves in both panels are obtained from numerical integration of Eqs. \ref{eq:model} with parameters $A_{11} = A_{22} = K = r = 1$, $\mu_S = 0.1$, and $A_{12} = 1.1$.}
    \label{fig3}
\end{figure}

\begin{figure*}
    \centering
    \includegraphics[width=2.0\columnwidth]{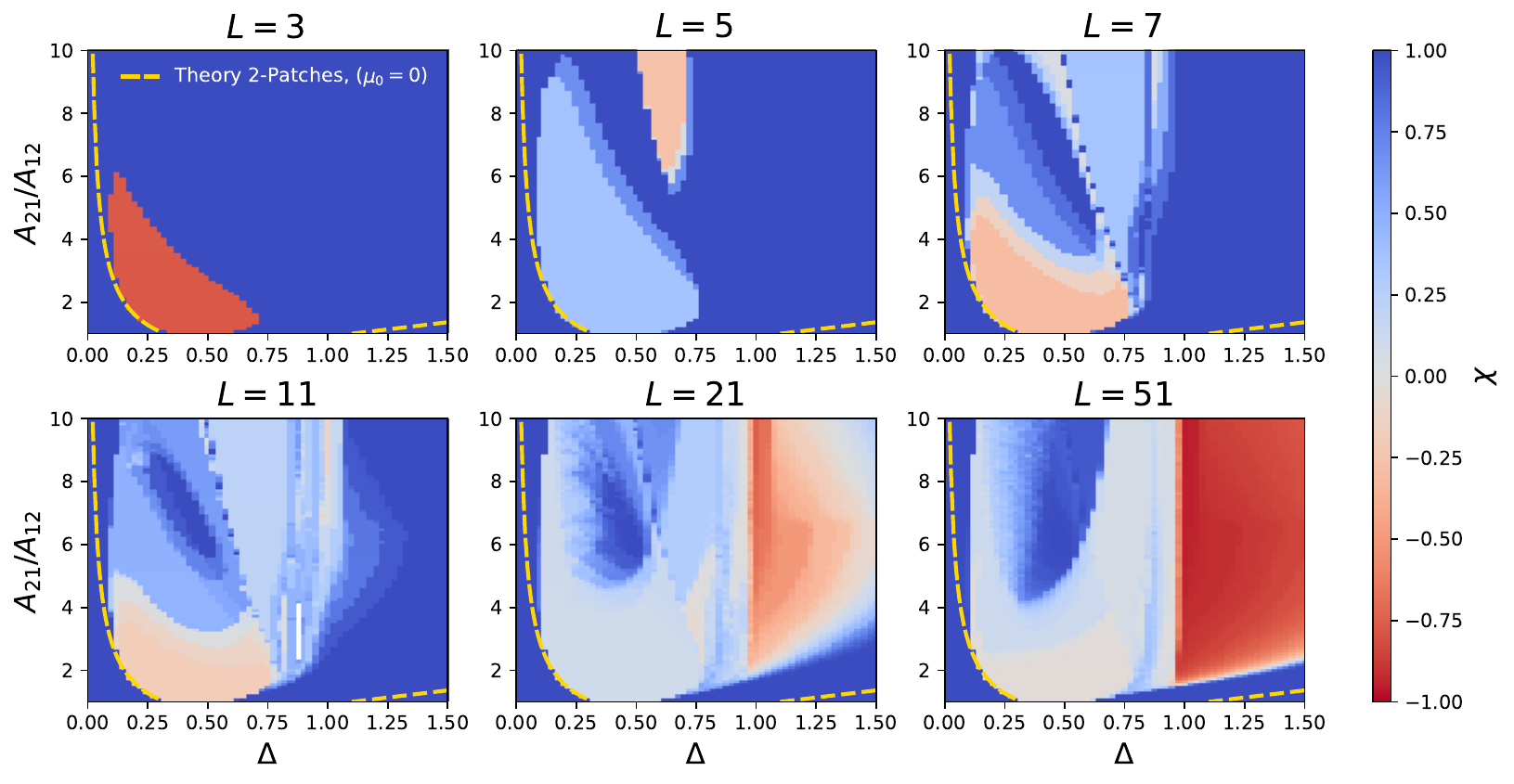}
    \caption{Heatmaps for lattices with different side lengths ($L = 3, 5, 7, 11, 21, 51$) showing the equilibrium difference in total abundances between species 1, $x_1=\sum_\alpha x_1^\alpha$, and species 2, $x_2=\sum_\alpha x_2^\alpha$. This difference is normalized by the total available area $L \times L$ and we call the normalized difference $\chi\equiv(x_1-x_2)/L^2$. Blueish regions indicate dominance of species 1, whereas reddish regions indicate dominance of species 2. As in figure \ref{fig2}, the activation threshold $\Delta$ is reported on the $x$-axis and the ratio of competition coefficients $A_{21}/A_{12}$ on the $y$-axis. The gold dashed line represents the analytical prediction obtained for the two-patch case without linear diffusion (see figure \ref{fig2}). Each cell in the heatmap corresponds to a numerical integration of eq. \ref{eq:model} with parameter values $A_{11} = A_{22} = K = r = 1$, $A_{12}=1.1$, $\xi=100$, $\mu_0 = 0.001$, and $\mu_S = 0.1$. Initial conditions are $x_1^\alpha(0) = x_2^\alpha(0) = 0.001$ in the central cell and zero elsewhere. For all lattices,  {\revDS the mobility matrices ${\bf M}$ are constructed considering a Moore neighborhood.}}
    \label{fig4}
\end{figure*}

We now relax the limits $\mu_0 \to 0$ and $\xi \to \infty$ to explore how increasing linear diffusion and reducing the steepness (and thus the nonlinearity in the competitor's abundance) of the motility response affect coexistence. In Fig. \ref{fig3} we show how the coexistence interval in motility thresholds, $\Delta_{\text{max}} - \Delta_{\text{min}}$, changes as a function of $\mu_0$ (Fig. \ref{fig3}a) and $\xi$ (Fig. \ref{fig3}b), for different values of the ratio $A_{21}/A_{12}$. Increasing linear diffusion monotonically shrinks the coexistence range, which eventually vanishes ($\Delta_{\text{max}} - \Delta_{\text{min}} = 0$) once $\mu_0$ is sufficiently large. Fig. \ref{fig3}a highlights the breakdown of the competition–colonization trade-off when linear diffusion becomes strong: for small diffusion, stronger asymmetry (larger $A_{21}/A_{12}$) enlarges the coexistence interval because it enhances the ability of species 2 to quickly escape and colonize patch $\beta$. In contrast, at higher diffusion rates, asymmetry no longer favors coexistence, as it mainly amplifies the exploitation of species 2 by species 1. Consequently, less asymmetric interactions require larger values of $\mu_0$ to eradicate coexistence.

Regarding the influence of the steepness of the motility response, we show in Fig. \ref{fig3}b that coexistence range exhibits a sharp increase with $\xi$ before saturating at the asymptotic value reached in the limit $\xi \to \infty$ for all values of $A_{21}/A_{12}$. This behavior indicates that a minimum degree of nonlinearity in the escape response is required to sustain coexistence. In Appendix B, we provide theoretical arguments showing that segregated coexistence cannot occur when $\xi \to 0$ in the regime $A_{12} > 1$ and $A_{21} > 1$.

In the next section, we extend this analysis to two-dimensional lattices of various sizes, showing that the simplified two-patch framework developed above remains to some extent informative in these more complex settings.}

\subsection{Two-dimensional Lattices}

{\revDS We now extend our analysis to more complex metapopulations beyond the two-patches system}. To this end, we consider multiple two-dimensional lattices composed of $L \times L$ cells, and we vary the side length $L$. As $L$ increases, this can be interpreted either as an expansion of the available space or, equivalently, as a reduction of the characteristic spatial scale at which bacteria interact and grow locally. The equations describing this extended setting are identical to those in eq. \ref{eq:model}, with $M_{\alpha\beta}=M_{\beta\alpha}=1/\mathcal{N}$ for neighboring cells, where $\mathcal{N}$ is the fixed number of neighbors of each cell, and $M_{\alpha\beta}=M_{\beta\alpha}=0$ otherwise.

As before, we initialize the system by placing a small seed of both species in a single cell, specifically the central one, to mimic experiments in which a central inoculum spreads across a Petri dish or a plate. {\revDS Likewise, we construct the mobility matrix ${\bf M}$ by assuming a Moore neighborhood, thus connecting each patch with each eight nearest neighbors in the lattice}. In the supplementary material we also analyze the case of more heterogeneous initial spatial distributions and {\revDS other mobility matrix by setting a Von Neumann neighborhood, connecting each patch with each four nearest neighbors}.  In the following analysis we fix $\xi = 100$, as we have seen that {\revDS the steepness of the sigmoid function} favors coexistence. We also keep a non-zero diffusion term fixed at $\mu_0 = 10^{-3}$. 

\begin{figure*}
    \centering
    \includegraphics[width=2.0\columnwidth]{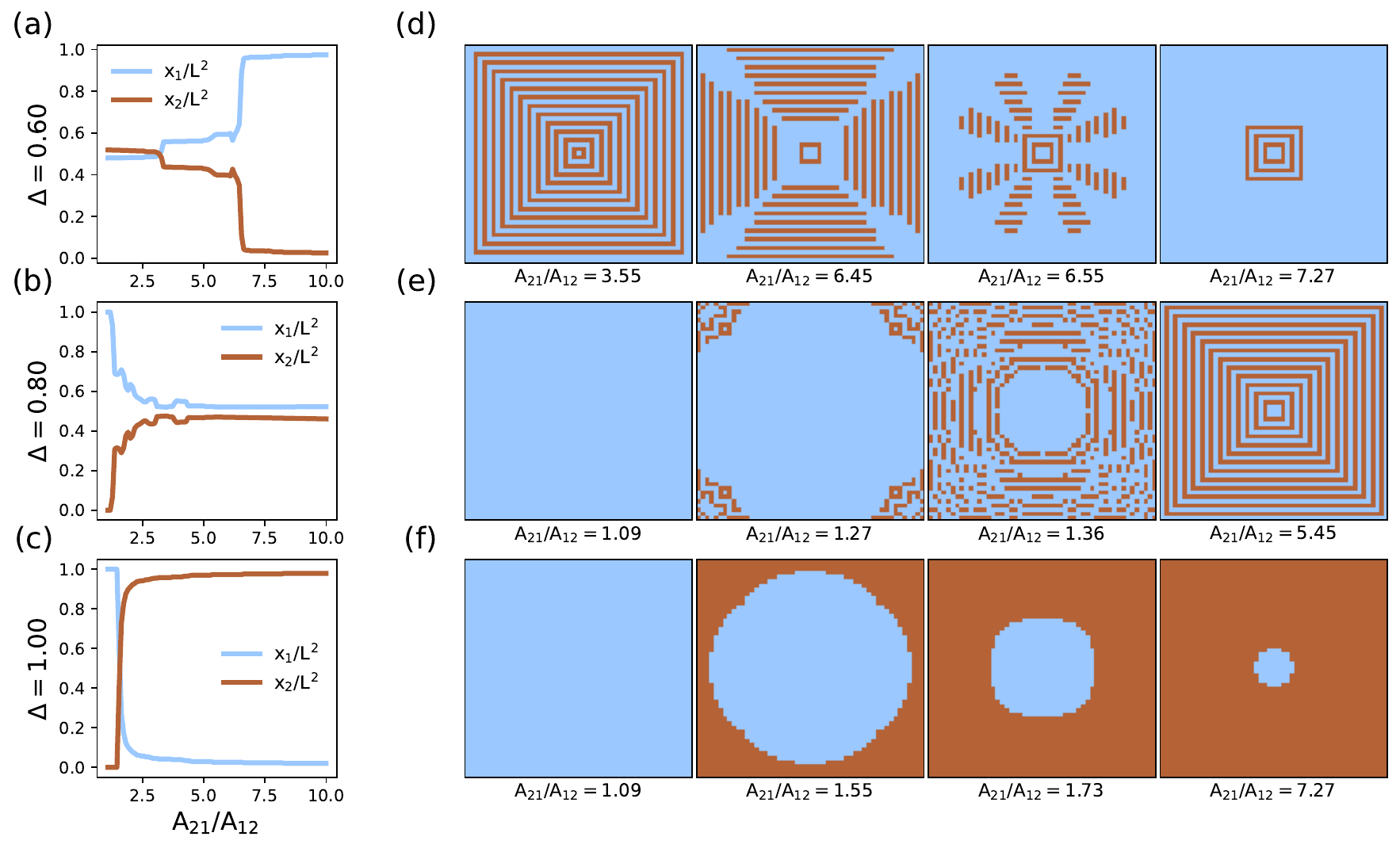}
    \caption{(a-c) Final total abundances of species 1 ($x_1$) and species 2 ($x_2$), normalized by the lattice size $L^2$, as a function of $A_{21}/A_{12}$, for three values of the activation thresholds $\Delta = 0.60,\ 0.80,\ 1.00$. (d-f) Stationary spatial patterns for species 1 (blue) and species 2 (red) on a two-dimensional lattice with side length $L = 51$. Rows correspond to activation thresholds $\Delta = 0.60,\ 0.80,\ 1.00$, and columns to increasing values of the competition ratio $A_{21}/A_{12}$.  All results are obtained from numerical integration of the model with parameters $A_{11} = A_{22} = K = r = 1$, $A_{12}=1.1$, $\xi=100$, $\mu_0=0.001$, and $\mu_S=0.1$. Initial conditions place both species at density $0.001$ in the central cell and zero elsewhere. {\revDS The mobility matrix ${\bf M}$ is constructed considering a Moore neighborhood.}}
    \label{fig5}
\end{figure*}

Figure \ref{fig4} displays the same phase diagram shown in Figure \ref{fig2} expressed in terms of the ratio of competition coefficients $A_{21}/A_{12}$ and the activation threshold $\Delta$, for multiple values of the spatial dimension $L$. In contrast to Figure \ref{fig2}, the heatmap here represents the total abundance difference $x_1-x_2 = \sum_{\alpha}(x_1^\alpha-x_2^\alpha)$, between the two species at equilibrium, normalized by the total available area $L^2$, that we indicate with $\chi$. The darkest blue region corresponds to parameter combinations for which coexistence is not achieved, and only the stronger species 1 persists. The most evident feature is that increasing $L$ leads to a general expansion of the coexistence region. {\alex Intuitively, larger lattices provide more empty frontier for the faster-escaping species to colonize before diffusive inflow can re-establish local competitive exclusion, thereby strengthening the competition--colonization trade-off.}
Within this region, the dominance of one species over the other depends on several factors, although some clear patterns emerge. For small values of $\Delta$ and increasing ratios $A_{21}/A_{12}$, the system consistently transitions from complete dominance of species 1 (non-coexistence) to coexistence with comparable abundances of the two species (white or lightly shaded blue/red areas). As $L$ increases, coexistence becomes possible over a broader range of $\Delta$; in particular, for large $L$ and $\Delta \ge 1$, a region appears in which the weaker competitor, species 2, becomes largely dominant. 

Remarkably, as the lattice size increases, the coexistence region approaches the phase diagram obtained in the two-patch system without standard diffusion ($\mu_0 = 0$), shown in Figure \ref{fig2} (the analytical prediction derived for that case is plotted as a dashed line in Figure \ref{fig4}). This convergence suggests that a system with limited spatial extent (or, equivalently, with a large interaction range) but no diffusion, behaves similarly to a larger spatial system (or shorter-range interactions) and nonzero diffusion. In this sense, the theoretical description developed for the simpler two-patch setting also captures to some extent the behavior of this more realistic spatial framework.

{\revDS 
To further examine these results, Figs.~\ref{fig5}a–c show how the equilibrium abundances of the two species vary with increasing interspecific competition for three different values of the motility threshold: $\Delta=0.6$ (Fig.~\ref{fig5}a), $\Delta=0.8$ (Fig.~\ref{fig5}b), and $\Delta=1.0$ (Fig.~\ref{fig5}c). These results illustrate how the interplay between Lotka–Volterra demographic dynamics and nonlinear motility responses leads to qualitatively different effects of increasing interspecific competition on coexistence. Depending on the value of $\Delta$, increasing the ratio $A_{21}/A_{12}$ can (i) reduce the abundance of the weaker species (Fig.~\ref{fig5}a), (ii) promote coexistence with comparable spatial abundances of the two species (Fig.~\ref{fig5}b), or (iii) even lead to the spatial dominance of the weaker competitor over the stronger one (Fig.~\ref{fig5}c).

To understand the disparate effect of interspecific competition, we explicitly study the spatial self-organization of the species by plotting representative spatial patterns observed at equilibrium for $\Delta=0.6$ (Fig.~\ref{fig5}d), $\Delta=0.8$ (Fig.~\ref{fig5}e), $\Delta=1$ (Fig.~\ref{fig5}f) and different values of $A_{21}/A_{12}$. In the coexistence regime with $\Delta < 1$, we consistently observe the emergence of concentric rings in which the two species alternate in space. This organization arises because, for small $\Delta$, the density-dependent motilities of both species eventually gets activated. Since $A_{12} < A_{21}$, species 2 reaches its activation threshold first and escapes outward, colonizing the first ring around the central cell, while species 1 diffuses more slowly with $\mu_0$. In this newly occupied ring, species 2 attains a density high enough to trigger the escape response of species 1, which then fills the next outer ring. This process repeats, producing alternating concentric domains of the two species.

As the competitive effect of species 1 on species 2 increases (i.e., for larger values of $A_{21}/A_{12}$), the concentric rings become destabilized and give rise to more irregular spatial patterns. This occurs because, under sufficiently strong competition, the small inflow of species 1 generated by the diffusion term $\mu_0$ can locally out-compete species 2 even where species 2 is initially more abundant (Fig.~\ref{fig5}d).

When $\Delta$ approaches 1, a different mechanism becomes relevant. For low values of $A_{21}/A_{12}$, the density-dependent motility of species 2 activates too late, and it may fail to survive. In this regime, stronger competition is required for species 2 to reach the activation threshold early enough to escape and re-establish the concentric rings pattern, as shown in Fig.~\ref{fig5}e. 

Finally, when $\Delta \ge 1$, the escape response of species 1 is hardly activated, and its spread is limited to standard diffusion. In contrast, species 2 escapes more rapidly as $A_{21}/A_{12}$ increases, colonizing regions farther from the center. Stronger competition leaves species 1 progressively less time to expand before species 2 establishes itself. This regime is shown in Fig.~\ref{fig5}f and it is particularly notable because species 2, despite being the inferior competitor, becomes substantially more abundant and occupies a larger spatial region than species 1. This counterintuitive behavior can be traced back to the colonization–exploitation trade-off, whereby enhanced motility allows the weaker species to preferentially occupy empty spatial patches.

}
\section{Discussion}

Microbial communities display both high taxonomic diversity and remarkable long-term stability in composition \cite{Thompson, HumanMicrobiomeProject}. This empirical evidence has motivated sustained theoretical efforts to extend classical ecological models, such as Lotka–Volterra and consumer–resource frameworks, which generally predict that increasing diversity leads to instability \cite{May}. One line of explanation reconciles diversity and stability by assuming weak or sparse interactions between species \cite{Grilli, Camacho}. An alternative view is that classical ecological frameworks are overly reductionist and lack key ingredients required to account for the observed levels of diversity. From this perspective, incorporating explicit spatial structure emerges as a promising direction \cite{Cordero}, as microbial communities are typically far from well mixed at local scales and instead exhibit pronounced patchiness and spatial segregation among species \cite{O'Brien, Sheth, Welch}. 

{\alex In this work, we show that, in a spatially structured model, coexistence among strongly competing species can be sustained by an escape response in which motility increases nonlinearly with the local abundance of competitors. In the competitive-exclusion regime ($A_{12}>1$ and $A_{21}>1$), coexistence does not occur within single patches, and therefore persistence at the metapopulation level relies on spatial self-organization.}
{\alex In our framework,} coexistence arises via a classical competition–colonization trade-off, {\alex which} emerges from coupling demographic {\alex interactions to the} sigmoidal, quorum-sensing-like functional form assumed for the escape response. In this setting, a high ability to escape from competitors also entails an increased susceptibility to being outcompeted, and coexistence relies on a balance between these opposing effects. If the inferior competitor grows too rapidly, it can trigger the escape of the superior competitor, ultimately leading to exclusion; conversely, if it grows too slowly, it risks being outcompeted, again resulting in exclusion. This interplay is what generates the rich phenomenology in the phase diagrams discussed previously.

The role of spatial segregation in promoting coexistence has a long history in theoretical ecology. In one of the first works in this direction, Levin \cite{Levin1974} showed that, in patchy environments, linear diffusive fluxes can have a stabilizing effect on the coexistence of competing species. In that framework, however, coexistence relied primarily on initial heterogeneity in patch colonization combined with sufficiently low diffusion rates. Subsequent work introduced nonlinearities in migration by allowing dispersal to depend explicitly on competitor densities, as in Shigesada \textit{et al.} \cite{Shigesada1979} and Mimura and Kawasaki \cite{Mimura1980}. In these models, the effect was typically implemented through terms proportional to the product of species abundances (e.g., $-c\,x_1^\alpha x_2^\alpha$, with $c$ a coupling constant). As in our case, these studies showed that nonlinear migration can enable coexistence in regimes where it is otherwise forbidden in well-mixed systems. 

{\revDS In all these aforementioned approaches}, growth, competition, and migration are typically decoupled, allowing coexistence to be recovered by independently tuning motility parameters. {\alex A key difference in our approach is that competition and motility are not tuned independently: the same interspecific interaction coefficients that determine demographic suppression also enter the escape response.}
{\alex Moreover, motility dependence on competitor abundance is strongly nonlinear}, being mediated by the sigmoid function. In the limit $\xi \to 0$, this formulation reduces to the linear (in terms of competitor's abundance) models discussed above and coexistence is no longer observed within our coupled setup {\alex (Fig.~\ref{fig3}b and Appendix~B).}
In our case, the weaker competitor is not endowed a priori with a higher colonization ability (especially when $A_{12}\sim A_{21})$. Instead, the competition–colonization trade-off emerges spontaneously from the growth dynamics: the stronger competitor grows faster and therefore activates the escape response of the weaker one earlier. This asymmetry arises solely from the nonlinear, threshold-like nature of the escape response, which is essential for the emergence of coexistence in our framework.

In ecology, near-perfect reciprocal exclusion of species across sites, leading to alternating patches, is commonly referred to as a checkerboard distribution \cite{Stone}. The origins of checkerboard distributions have long been debated \cite{Dallas}. While they are often interpreted as signatures of interspecific competition \cite{Diamond}, it has also been argued that similar patterns may arise purely from environmental filtering, even in the absence of direct interactions \cite{Connor}. {\alex Our results identify an additional mechanism through which competition, when coupled to a strongly nonlinear escape response, can both generate and maintain checkerboard-like spatial organization.}

Beyond coexistence, {\revDS we have also shown the emergence of rich spatial pattern formation in lattice geometries as a function of the motility threshold and the inter-specific competition}. 
{\alex Some of them,}
specifically the concentric ring structures arising from the reciprocal activation of motility in two competing species, have been previously observed both in silico and experimentally by Curatolo \textit{et al.} \cite{Curatolo}. 
{\alex Classical active-matter frameworks used to describe MIPS-like behavior often rely on coarse-grained continuum equations for density fields; here, instead, we adopt a discrete metapopulation framework that is standard in theoretical ecology when the primary focus is on coexistence and persistence \cite{Gravel, Denk, Lorenzana}.}
As a representation of continuous space, metapopulations models have limitations: {\alex within each patch, populations are assumed locally well mixed, and demographic interactions occur at a mesoscopic scale set by the patch size. Systematically refining the spatial description corresponds, in our setting, to increasing the number of patches (increasing $L$) and thereby reducing the local interaction scale.}

{\revDS In conclusion, the theoretical framework here introduced represents one of the first attempts to frame the exploitation-colonization trade-off as the natural coupling between demographic interactions and motility responses driven by quorum-sensing like mechanisms. In a broader perspective, our findings are relevant to two closely related yet distinct research directions: the identification of mechanisms that promote coexistence in microbial ecology, and the introduction of a spatial framework that generates novel patterns of relevance to active-particle physics.}

\begin{acknowledgments}
We acknowledge support from Spanish Ministerio de Ciencia e Innovacion (PID2024-158120NB-C21), Generalitat de Catalunya (2021SGR-00633) and ICREA Academia. This project has received funding from the European Union's Horizon 2020 research and innovation programme under the Marie Skłodowska-Curie grant agreement No. 945413 and from the Universitat Rovira i Virgili (URV).
\end{acknowledgments}

\section*{Authors Contribution}

M.M., D.S.-P. and A.A. designed the study; M.M. performed analysis,  mathematical calculations, and wrote code; all the authors wrote the paper and discussed the results.

\appendix
\section{Analytical Description for the Two-Patch Setting ($\xi\to\infty$, $\mu_0\to 0$)}

We recall the two-patch scenario in which two species, labeled 1 and 2, inhabit two connected but distinct patches, $\alpha$ and $\beta$, whose dynamics follow Eq.~(\ref{eq:model}). Throughout this appendix we consider the limits $\xi \to \infty$ and $\mu_0 \to 0$, so that the motility function reduces to a step function. As already mentioned in the main text, we start from an initial condition where a small seed of both species is present only in patch $\alpha$, while patch $\beta$ is initially empty.

In this regime, stable coexistence can occur only if the inferior competitor, here species 2, activates the escape response and colonizes patch $\beta$, while species 1 does not. We recall that this requirement is equivalent to imposing the conditions established Eqs.~(\ref{eq:cond1})-(\ref{eq:cond2}). The second condition is automatically satisfied if $A_{21} < \Delta$, since species 1 always dominates locally and reaches the carrying capacity $K=1$. To determine when the first condition holds, we must compute explicitly the time evolution of $x_2^\alpha(t)$ for all $t$. In order to do so, we approximate the equation for species 1 in patch $\alpha$ as
\begin{equation}
\frac{dx_1^\alpha(t)}{dt}
= x_1^\alpha(t)\,\big(1 - x_1^\alpha(t) - A_{12}\,\epsilon \big),
\label{eq:A1}
\end{equation}
where we have replaced $x_2^\alpha(t)$ with a small constant value $\epsilon$. This approximation is valid when species 2 activates the escape response sufficiently early, so that species 1 experiences only weak competitive pressure arising from the low abundance of species 2. With this simplification and rewriting 
\begin{equation*}
\begin{split}
\frac{1}{x_1^\alpha(t)\,(1-x_1^\alpha(t)-A_{12}\,\epsilon)} =\frac{1}{1-\epsilon}\times\\
\Bigg(\frac{1}{x_1^\alpha(t)}\,\,+\,\,
\frac{1}{1-x_1^\alpha(t)-A_{12}\,\epsilon}\Bigg),
\end{split}
\end{equation*}
we can easily integrate equation \ref{eq:A1}, yielding
\begin{equation}
x_1^\alpha(t)
=
\frac{
    (1 - A_{12}\,\epsilon)
}{
    1
    + \bigg(
        \dfrac{1 - x_1^\alpha(0) - A_{12}\,\epsilon}{x_1^\alpha(0)}
      \bigg)
      e^{-t(1 - A_{12}\,\epsilon)}
},
\end{equation}
which is a shifted logistic form. With this we can now express explicity the time $\tau$ at which $x_1^\alpha(\tau)=\Delta/A_{21}$, i.e. the time at which the escape response is activated and species 2 starts to migrate towards patch $\beta$. After some algebra
\begin{equation}
\begin{aligned}
    x_1^\alpha(\tau) &= \frac{\Delta}{A_{21}}
    \iff 
    \tau = \frac{1}{\,1 - A_{12}\,\epsilon\,} \times\\
    &\qquad
    \ln\!\bigg[
        \frac{\Delta}{A_{21}}
        \frac{
            (1 - x_1^\alpha(0) - A_{12}\,\epsilon)/x_1^\alpha(0)
        }{
            1 - A_{12}\,\epsilon - \Delta/A_{21}
        }
    \bigg].
\label{eq:tau}
\end{aligned}
\end{equation}

For $t<\tau$, the variable $x_2^\alpha(t)$ experiences neither in-migration nor out-migration and therefore evolves solely according to the Lotka--Volterra dynamics
\begin{equation}
    \frac{dx_2^\alpha(t<\tau)}{dt}
    =
    x_2^\alpha(t)\,\Big(1 - x_2^\alpha(t) - A_{21}\,x_1^\alpha(t)\Big).
\end{equation}
Since the coupling between $x_1^\alpha(t)$ and $x_2^\alpha(t)$ has been removed in {\revDS Eq.~(\ref{eq:A1})}, this equation becomes a quadratic Bernoulli differential equation, which can be solved explicitly. By defining
\begin{equation}
    E(t)\equiv \exp\bigg[\int_0^t\big(1 - A_{21}x_1^\alpha(t')\big)\,dt'\bigg],
\end{equation}
and applying a simple change of variables, one finds that the solution can be written as
\begin{equation}
    x_2^\alpha(t<\tau)
    =
    \frac{E(t)}{\,1/x_2^\alpha(0) + \int_0^t E(t')\,dt'}.
\label{eq:x2a}
\end{equation}

In our setup, both $E(t)$ and its integral can be computed explicitly.  
The first one, after a simple change of variables, reduces to an elementary integral and yields
\begin{equation}
    E(t)
    = e^{t}\,\Bigg(
        \frac{x_0\big(e^{t(1-A_{12}\,\epsilon)} - 1\big) - A_{12}\,\epsilon + 1}{\,1 - A_{12}\,\epsilon}
      \Bigg)^{-A_{21}},
\label{eq:E}
\end{equation}
where, for simplicity, we have introduced the shorthand notation  
$x_1^\alpha(0) = x_2^\alpha(0) \equiv x_0$. The time integral of the latter can be written as
\begin{equation}
\begin{aligned}
    \int_0^t E(t')\,dt'
    &= (1 - A_{12}\,\epsilon)^{A_{21}}\,\,\times \\
    &\quad
    \int_0^t
    \frac{
        e^{t'}
    }{
        \Big[x_0\big(e^{t'(1 - A_{12}\,\epsilon)} - 1\big) - A_{12}\,\epsilon + 1\Big]^{A_{21}}
    }\,dt'.
\end{aligned}
\end{equation}
which has a known solution in terms of incomplete beta functions $B_x(a,b)$%
\cite{nisturl}:
\begin{equation}
\begin{split}
    \int_0^t E(t')\,dt'
    = (1-A_{12}\,\epsilon - x_0)^{\frac{1 - A_{21} + A_{12}A_{21}\epsilon}{1 - A_{12}\epsilon}}
    \times \\
    (1-A_{12}\,\epsilon)^{A_{21}-1}\,
    x_0^{-\frac{1}{1-A_{12}\epsilon}}
    \times\\
    \Bigg[B_{X_{t'}}\!\bigg(
        \frac{1}{1-A_{12}\,\epsilon},
        \frac{A_{21} - \epsilon\,A_{12}A_{21} - 1}{1 - A_{12}\,\epsilon}
    \bigg)\Bigg]_{t'=0}^{t'=t},
\label{eq:intE}
\end{split}
\end{equation}
where
\begin{equation}
    X_{t'} \equiv
    \frac{x_0\, e^{(1-A_{12}\epsilon)t'}}
         {1 - A_{12}\epsilon - x_0\big(1 - e^{(1-A_{12}\epsilon)t'}\big)}.
\end{equation}
Substituting expressions \ref{eq:E} and \ref{eq:intE} into equation \ref{eq:x2a}, we obtain a complete description of $x_2^\alpha(t)$ for $t<\tau$. At $t=\tau$, the dynamics of $x_2^\alpha(t)$ changes due to the appearance of a migration term, $-\mu_S\,x_2^\alpha(t)$, representing the flux from patch $\alpha$ toward patch $\beta$. Therefore, for $t>\tau$, the evolution is given by
\begin{equation}
    x_2^\alpha(t>\tau)
    = x_2^\alpha(t)\Big(1 - x_2^\alpha(t) - A_{21}x_1^\alpha(t) - 2\mu_S\,x_2^\alpha(t)\Big),
\end{equation}
which is again a Bernoulli equation, and its solution has the same structure as in \ref{eq:x2a}. The procedure is identical to the previous case, but with a modified integrating factor $E'(t)$ and its corresponding integral, both affected by the migration term and by the fact that integration now starts at $t=\tau$. Explicitly,
\begin{equation}
\begin{split}
    E'(t)
    &= e^{(t-\tau)(1-2\mu_S)} \\
    &\quad\times
    \Bigg(
        \frac{
            e^{t(1-A_{12}\epsilon)} + (1-x_0-\epsilon)/x_0
        }{
            e^{\tau(1-A_{12}\epsilon)} + (1-x_0-\epsilon)/x_0
        }
    \Bigg)^{-A_{21}} .
\end{split}
\end{equation}

and
\begin{equation}
\begin{split}
    \int_\tau^t E'(t')\,dt'
    = \frac{
        e^{\tau(2\mu_S-1)}
    }{
        \Big(
            e^{\tau(1-A_{12}\,\epsilon)} + (1-x_0-A_{12}\,\epsilon)/x_0
        \Big)^{-A_{21}}
    }
     \\
    \times\frac{1}{1-A_{12}\,\epsilon}
    \left(
        \frac{1-A_{12}\,\epsilon - x_0}{x_0}
    \right)^{\frac{1 - 2\mu_S - A_{21} + \epsilon A_{12}A_{21}}{1-A_{12}\,\epsilon}}
     \\
    \times\Bigg[
        B_{X'_{t'}}\!\left(
            \frac{1-2\mu_S}{1-A_{12}\,\epsilon},
            \frac{A_{21} - \epsilon A_{12}A_{21} + 2\mu_S - 1}{1-A_{12}\,\epsilon}
        \right)
    \Bigg]_{t'=\tau}^{t'=t},
\end{split}
\end{equation}
with
\begin{equation}
    X'_{t'}\equiv\frac{e^{(1-A_{12}\,\epsilon)\,t'}}{e^{(1-A_{12}\,\epsilon)\,t'}+\frac{1-A_{12}\,\epsilon-x_0}{x_0}}.
\end{equation}
With $E'(t)$ and $\int_\tau^t E(t')\,dt'$, we can compute $x_2^\alpha(t)$ for $t \ge \tau$ again as
\begin{equation}
    x_2^\alpha(t \ge \tau)
    =
    \frac{E'(t)}{\,1/x_2^\alpha(\tau) + \int_\tau^t E(t')\,dt'},
\label{eq:x2a_2}
\end{equation}
where $x_2^\alpha(\tau)$ is obtained from the previous phase (eq. \ref{eq:x2a}), together with the explicit formula for $\tau$ (eq. \ref{eq:tau}).

With these expressions at hand, it is possible to determine which combinations of parameters lead to coexistence through segregation. Indeed, the theoretical curves shown in figures \ref{fig2} and \ref{fig4} are obtained by computing the minimum value of $\Delta$ for which condition \ref{eq:cond1}, namely
\[
    x_2^\alpha(t) < \Delta/A_{12}, \qquad \forall t,
\]
is satisfied, using $x_2^\alpha(t)$ from eq. \ref{eq:x2a_2}. All the curves depicted in the figures in the main text are obtained considering $\epsilon=0.05$.

\section{The limit $\xi \to 0$}

Here we analyze the limit in which the sigmoidal escape response becomes linear in the competitor abundance, corresponding to $\xi \to 0$.  
For small argument $x$, the sigmoid admits the expansion
\begin{equation}
    \frac{1}{1+e^x}
    =
    \frac{1}{2}
    -
    \frac{x}{4}
    +
    \mathcal{O}\!\left(x^3\right).
\end{equation}
Applying this expansion, for instance, to the motility rate of species 1 from patch $\alpha$ to $\beta$, we obtain to leading order
\begin{equation}
    \mu_1^{\alpha\beta}(t)
    \simeq
    M_{\alpha\beta}
    \bigg[
        \mu_0
        +
        2\mu_S
        \bigg(
            \frac{1}{2}
            -
            \frac{\xi\Delta}{4}
            +
            \frac{\xi A_{12}\, x_2^\alpha(t)}{4}
        \bigg)
    \bigg].
\end{equation}

With this approximation, the system of four equations describing the two-patch model becomes analytically tractable. We investigate whether segregated coexistence remains possible in this linearized limit, i.e., whether a steady state of the form
\[
\bigl(x_1^\alpha, x_1^\beta, x_2^\alpha, x_2^\beta\bigr)
=
\bigl(1-\epsilon_1,\, \epsilon_1,\, \epsilon_2,\, 1-\epsilon_2\bigr),
\qquad
\epsilon_1,\epsilon_2 \ll 1,
\]
can exist.

Linearizing the steady-state equations and solving those for $x_1^\alpha$ and $x_1^\beta$, we obtain
\begin{equation}
    \epsilon_1
    =
    \frac{
        \mu_S(2-\xi\Delta)
    }{
        2\big[
            \mu_S(2-\xi\Delta)
            +
            (\xi\mu_S+1)(A_{12}-1)
        \big]
    },
\label{eq:eps1}
\end{equation}
and
\begin{equation}
\begin{split}
    \epsilon_2
    &=
    \frac{
        \mu_S(2-\xi\Delta)(2-A_{12})
    }{
        2A_{12}\big[
            \mu_S(2-\xi\Delta)
            +
            (\xi\mu_S+1)(A_{12}-1)
        \big]
    }
    \\
    &=
    \left(\frac{2}{A_{12}}-1\right)\epsilon_1,
\end{split}
\label{eq:eps2}
\end{equation}
where, for readability, we have taken the limit $\mu_0 \to 0$.

Before enforcing the remaining steady-state equations for $x_2$, which we will refer to as $E_3$ and $E_4$, two trivial cases leading to full segregation are already apparent. The first corresponds to $\mu_S=0$, for which no motility is present and segregation can only persist if it is imposed by the initial conditions. The second occurs when $\xi\Delta=2$, which in the limit $\xi\to0$ requires $\Delta\to\infty$. Moreover, even for finite $\xi$ and large but finite $\Delta$, the equations $E_3$ and $E_4$ do not vanish when $\epsilon_1=\epsilon_2=0$, since $\mu_2^{\alpha\beta}(t)\neq0$.

We therefore focus on the nontrivial case $\epsilon_1>0$ and $\epsilon_2>0$. From Eqs. \eqref{eq:eps1}–\eqref{eq:eps2}, this requires
\begin{equation}
\begin{aligned}
\begin{cases}
2 - \xi\Delta > 0, \\
A_{12} < 2,
\end{cases}
\quad \text{or} \quad
\begin{cases}
2 - \xi\Delta
<
\dfrac{(\xi\mu_S+1)(A_{12}-1)}{\mu_S}, \\
A_{12} < 2.
\end{cases}
\end{aligned}
\end{equation}

Substituting Eqs. \eqref{eq:eps1} and \eqref{eq:eps2} into the linearized equations $E_3$ and $E_4$, we obtain
\begin{equation}
    E_3=-\epsilon_1 P,
    \qquad
    E_4=-\epsilon_1\bigl[P+2(A_{12}+A_{21}-2)\bigr],
\end{equation}
with
\begin{equation}
\begin{aligned}
P
=\;&
(A_{12}+A_{21}-2)
\Bigl[
A_{12}
+
\xi\mu_0(1-\Delta)(A_{12}-1)
\Bigr]
\\
&+
(A_{12}-1)
\Bigl[
A_{12}
+
\xi\mu_0(1-\Delta)
\Bigr].
\end{aligned}
\end{equation}

Both $E_3=0$ and $E_4=0$ can only be satisfied if $A_{12}+A_{21}=2$, which lies outside our competitive exclusion regime ($A_{12}>1$, $A_{21}>1$). We therefore conclude that, within our coupled framework, the limit $\xi\to0$ does not permit segregated coexistence. This result confirms the numerical findings shown in panel (b) of figure \ref{fig3} and demonstrates that a strongly nonlinear, threshold-like escape response is a necessary ingredient for coexistence.


\bibliography{bibliography} 

\clearpage
\onecolumngrid

\setcounter{figure}{0}
\renewcommand{\thefigure}{S\arabic{figure}}

\section*{Supplementary Material}

In this Supplementary Material, we examine how the phase diagram and the regions of coexistence and exclusion are affected by variations in the model setup. In particular, we explore the impact of different initial conditions, alternative neighborhood definitions, and changes in parameters that were fixed in the main text for simplicity. All results presented here correspond to a two-dimensional lattice with side length $L=51$, which we take as a representative case.

\subsection{Heterogeneous initial spatial distributions}

In the main text, we have focused on the case in which a small seed of both species is initially placed only in the central cell of the lattice. Under this choice, there is no intrinsic asymmetry between species, and the subsequent dynamics are entirely determined by how the two populations colonize the surrounding empty space. Here, we instead investigate a different class of initial conditions, in which both species are seeded in every cell of the lattice. A perfectly homogeneous initial distribution would trivially prevent coexistence, as the superior competitor would have an advantage in essentially every cell. To avoid this, we sample the initial abundances of both species in each cell independently from a uniform distribution, $x_1^\alpha(0),\,x_2^\alpha(0)\sim U[0,x_0]$, with $x_0$ small. This procedure generates strongly heterogeneous initial conditions across space. We then examine whether such heterogeneity is sufficient to sustain coexistence, identify the regimes in which it fails, and characterize the spatial patterns that emerge.

\begin{figure}
    \centering
    \includegraphics[width=0.8\columnwidth]{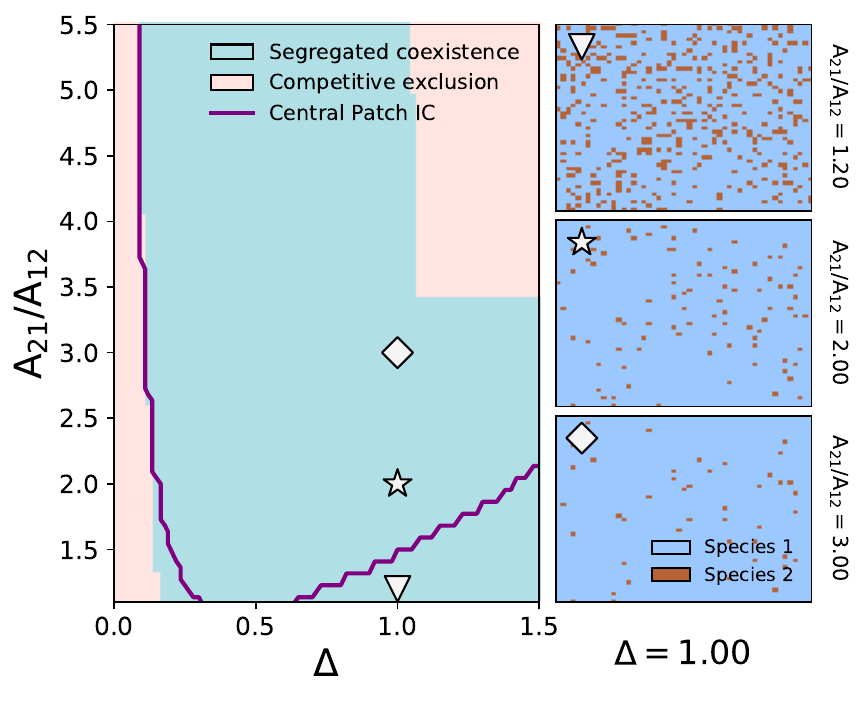}
    \caption{Phase diagram distinguishing segregated coexistence from competitive exclusion for heterogeneous initial spatial distributions of the two species. Initial abundances in each lattice cell are independently sampled from a uniform distribution, $x_1^\alpha(0),\,x_2^\alpha(0)\sim U[0,x_0]$, with $x_0=0.001$. The activation threshold $\Delta$ is shown on the $x$-axis, while the ratio of competition coefficients $A_{21}/A_{12}$ is reported on the $y$-axis. The solid purple line indicates the boundary between coexistence and exclusion obtained for the case in which both species are initially seeded only in the central cell (as in figure 4 of the main text). Representative stationary spatial distributions for $\Delta=1.0$ and $A_{21}/A_{12}=1.2,\,2,\,3$ are shown and indicated by the white markers. All results are obtained from numerical integration of the model with parameters $A_{11} = A_{22} = K = r = 1$, $A_{12}=1.1$, $\mu_S=0.1$, $\mu_0=0.001$, and $\xi=100$, on a two-dimensional lattice of side length $L=51$.}
    \label{figS1}
\end{figure}

In Figure \ref{figS1} we report the same phase diagram shown in Figure 2 of the main text, highlighting regions of coexistence at equilibrium versus exclusion as a function of the activation threshold $\Delta$ and the competition ratio $A_{21}/A_{12}$. For comparison, we also overlay the contour lines corresponding to the case in which the initial seeds are placed only in the central patch, as in figure 4 of the main text. Remarkably, for small values of $\Delta$ we observe a transition from exclusion to coexistence that is very similar to that obtained with central seeding. This indicates that, in this regime, the conditions required for the inferior competitor to escape, derived in Appendix A, depend only weakly on the choice of initial conditions. Instead, they are primarily controlled by the strength of competition between the species and by the critical competitor density at which the escape response is activated. These mechanisms are sufficient to promote coexistence, as the strong heterogeneity of the initial conditions  will inevitably create somewhere regions in which the inferior competitor is locally more abundant and can dominate. 

By contrast, for larger values of $\Delta$, the transition between exclusion and coexistence observed in the central-seed case is no longer present. Instead, we observe a transition from coexistence to exclusion as the competition ratio $A_{21}/A_{12}$ increases. In this regime, species 1 primarily undergoes linear diffusion and rarely, if ever, activates its escape response, whereas species 2 does. Under central-seed initial conditions, species 2 can exploit the competition–colonization trade-off to rapidly occupy empty space relative to the more slowly diffusing species 1. In the present case, however, the amount of available empty space is greatly reduced. As a result, species 2 can successfully spread only when competition is more balanced, allowing it to locally outcompete species 1 when the latter is sufficiently rare. As the asymmetry in competitive interactions increases, progressively smaller local abundances of species 1 are sufficient to suppress the escape-driven advantage of species 2, ultimately leading to the exclusion of species 1.

In this case, the variety of spatial patterns is markedly reduced compared to the central-seed scenario. The system typically settles into configurations characterized by isolated clusters of the inferior competitor embedded within extended regions dominated by the superior species. As the competition asymmetry increases (i.e., for larger $A_{21}/A_{12}$), these clusters become progressively smaller and less numerous, eventually disappearing (figure \ref{figS1}).

\subsection{Von-Neumann Neighborhood}

Here we consider the case in which bacterial motion is restricted to four directions (up, down, left, and right), rather than the eight directions allowed in the main text. This corresponds to adopting a von Neumann neighborhood instead of a Moore neighborhood.  

\begin{figure}
    \centering
    \includegraphics[width=1\columnwidth]{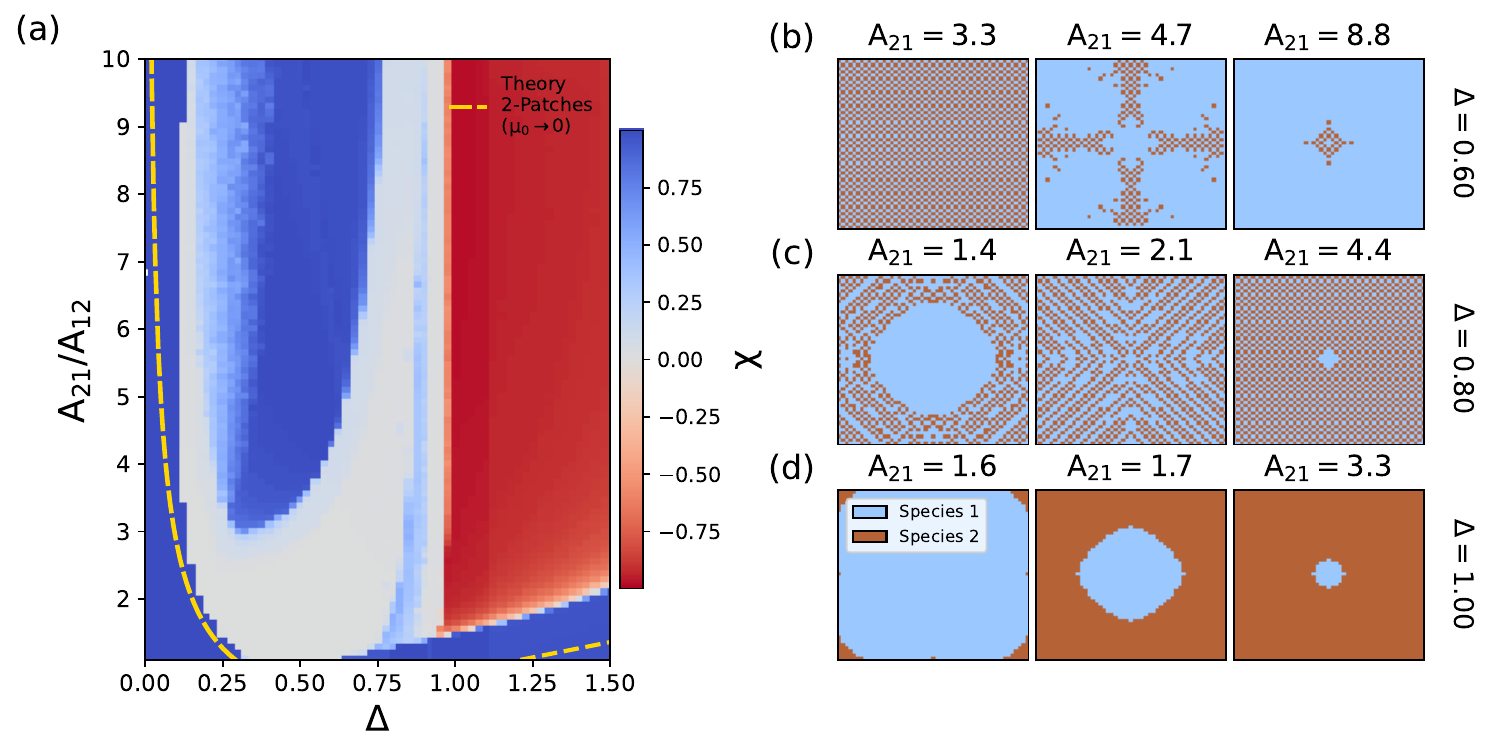}
    \caption{(a) Heatmap of the normalized difference between the total equilibrium abundances of species~1, $x_1=\sum_\alpha x_1^\alpha$, and species~2, $x_2=\sum_\alpha x_2^\alpha$. The quantity shown is $\chi\equiv(x_1-x_2)/L^2$, obtained by normalizing the abundance difference by the total system area $L^2$. Blue tones correspond to regimes in which species~1 is more abundant, while red tones indicate dominance of species~2. The horizontal axis reports the activation threshold $\Delta$, and the vertical axis shows the ratio of competition coefficients $A_{21}/A_{12}$. The gold dashed curve denotes the analytical boundary predicted by the two-patch theory in the absence of linear diffusion.  (b–d) Equilibrium spatial configurations for three representative values of the activation threshold, $\Delta=0.6$, $0.8$, and $1.0$. Cells dominated by species~1 are shown in blue, whereas cells dominated by species~2 are shown in red. Within each panel, three increasing values of $A_{21}$ are displayed: $A_{21}=3.3,\,4.7,\,8.8$ in (b); $A_{21}=0.4,\,2.1,\,4.4$ in (c); and $A_{21}=1.6,\,1.7,\,3.3$ in (d), chosen to best illustrate the qualitative dependence of spatial organization on the model parameters.  All results are obtained from numerical integration of the model with parameters $A_{11}=A_{22}=K=r=1$, $A_{12}=1.1$, $\xi=100$, $\mu_0=0.001$, $\mu_S=0.1$, and lattice side length $L=51$. Initial conditions correspond to $x_1^\alpha(0)=x_2^\alpha(0)=0.001$ in the central cell and zero elsewhere. The mobility matrices ${\bf M}$ are constructed using a von Neumann neighborhood.
}
    \label{figS2}
\end{figure}

Under this modification, we observe no substantial qualitative differences with respect to the results obtained for the Moore neighborhood. As in the main text, panel (a) of Fig. \ref{figS2} shows the phase diagram as a function of the activation threshold $\Delta$ and the competition ratio $A_{21}/A_{12}$, with the normalized difference in total abundances at equilibrium, $\chi=\sum_\alpha (x_1^\alpha-x_2^\alpha)/L^2$, used as the plotted quantity, as in Fig. 4 of the main text. The gold dashed line again denotes the analytical boundary of the coexistence region predicted by the two-patch theory in the absence of linear diffusion. The resulting phase diagram is practically almost identical to that obtained for the Moore neighborhood.

The only appreciable differences arise in the spatial organization of the populations, shown in panels (b–d) of Fig. \ref{figS2}. In particular, the concentric ring structures observed with the Moore neighborhood are replaced by perfectly alternating checkerboard patterns, as a direct consequence of the underlying neighborhood geometry. Despite this change in spatial morphology, the qualitative dependence of the patterns on $\Delta$ and $A_{21}/A_{12}$ remains the same as in the Moore case. For smaller values of $\Delta$, increasing competition asymmetry leads to the progressive dismantling of the checkerboard structure (panel (b)); for larger $\Delta$, checkerboard patterns are gradually stabilized as $A_{21}/A_{12}$ increases (panel (c)); and for $\Delta \geq 1$, dominance of the weaker competitor again emerges, as in the main text (panel (d)).

Overall, these results demonstrate that the specific choice of neighborhood rule in the discrete model does not qualitatively affect the main conclusions. While we believe that the Moore neighborhood provides a closer approximation to continuous space (this is confirmed by the experimental observations of concentric ring patterns in Curatolo et al. \cite{Curatolo}), checkerboard-like arrangements are themselves well documented in ecological systems \cite{Stone, Dallas, Diamond, Connor}, as already discussed in the main text. 

\subsection{Sensitivity Analysis}

Rather than modifying the structural setup of the model, we now examine the role of parameters that were held fixed in the main text. In particular, we focus on the strength of the competitive pressure exerted by species~2 on species~1, quantified by $A_{12}$, as well as on the motility rates, including both the linear diffusion coefficient $\mu_0$ and the amplitude $\mu_S$ associated with the sigmoidal escape response.

\subsubsection{Varying Competitive Pressure $A_{12}$}

In the main text, we fixed the interspecific competition coefficient $A_{12}=1.1$, an arbitrary choice slightly above unity, corresponding to the onset of competitive exclusion, and focused exclusively on increasing the asymmetry between competitors by varying $A_{21}$. However, as we show here, the system’s behavior is not determined solely by the ratio $A_{21}/A_{12}$: the absolute values of the competition coefficients $A_{12}$ and $A_{21}$ can also qualitatively affect the dynamics and the resulting phase diagram.

\begin{figure}
    \centering
    \includegraphics[width=1\columnwidth]{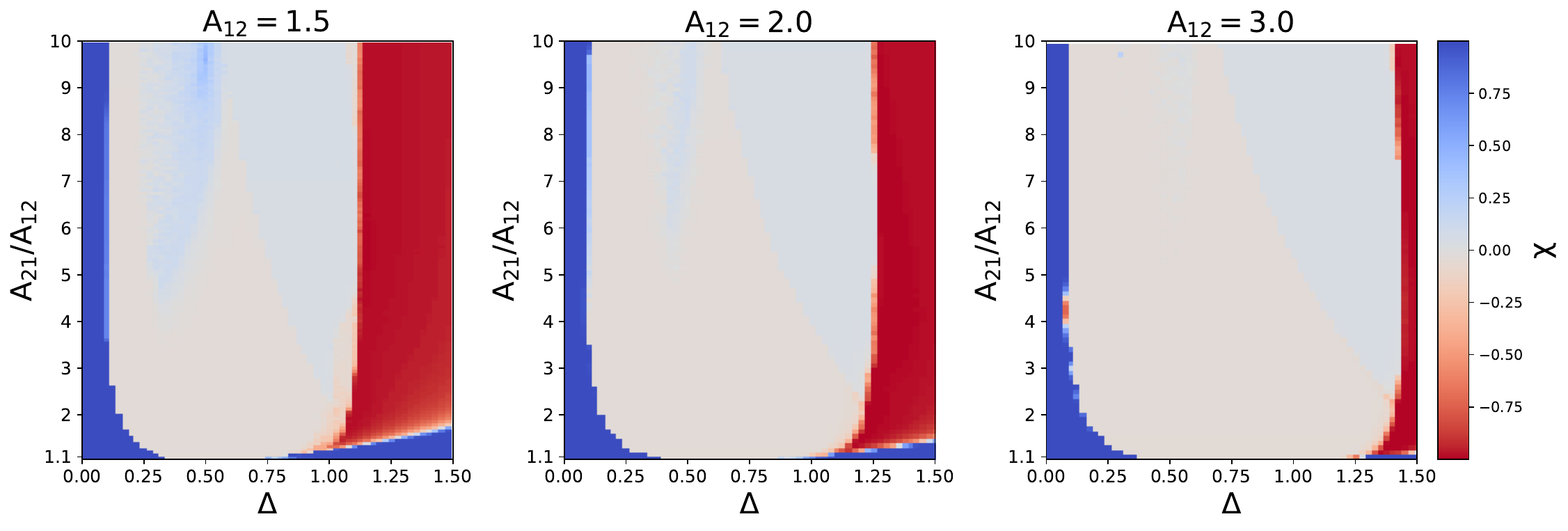}
    \caption{Heatmaps for different values of $A_{12}=1.5,\,2.0,\,3.0$ of the normalized difference between the total equilibrium abundances of species~1, $x_1=\sum_\alpha x_1^\alpha$, and species~2, $x_2=\sum_\alpha x_2^\alpha$. The quantity shown is $\chi\equiv(x_1-x_2)/L^2$, obtained by normalizing the abundance difference by the total system area $L^2$. Blue tones correspond to regimes in which species~1 is more abundant, while red tones indicate dominance of species~2. White/grey areas indicate similar abundances between the two species. The horizontal axis reports the activation threshold $\Delta$, and the vertical axis shows the ratio of competition coefficients $A_{21}/A_{12}$. All results are obtained from numerical integration of the model with parameters $A_{11}=A_{22}=K=r=1$, $\xi=100$, $\mu_0=0.001$, $\mu_S=0.1$, and lattice side length $L=51$. Initial conditions correspond to $x_1^\alpha(0)=x_2^\alpha(0)=0.001$ in the central cell and zero elsewhere. The mobility matrices ${\bf M}$ are constructed using a Moore neighborhood.}
    \label{figS3}
\end{figure}

Figure~\ref{figS3} illustrates how the phase diagram is modified as $A_{12}$ is increased. Specifically, we consider $A_{12}=1.5$, $2.0$, and $3.0$, and for each case vary $A_{21}$ up to $A_{21}/A_{12}=10$, as in the main text.
The primary effect of increasing $A_{12}$ is a systematic enlargement of the region in parameter space where concentric ring patterns of alternating species~1 and species~2 emerge. Indeed, in Fig.~\ref{figS3} all three heatmaps display an extended grey region, indicating comparable total abundances of the two species. In this parameter regime, such balance can only be achieved through spatial segregation into concentric rings. Compared to the $A_{12}=1.1$ case discussed in the main text, increasing $A_{12}$ progressively suppresses the bluish regions observed at intermediate $\Delta$s and large $A_{21}/A_{12}$, where concentric rings become unstable and the stronger competitor regains dominance (see figures 4 and 5 in the main text). This occurs because, even though the same range of $A_{21}/A_{12}$ is explored, increasing $A_{12}$ also increases the ability of the weaker competitor to invade new patches and outcompete small seeds of the stronger competitor. As a result, the alternating ring structures are stabilized over a broader parameter range.

For each value of $A_{12}$, we still observe a regime at large $\Delta$ in which the weaker competitor dominates in space. Increasing $A_{12}$ shifts the boundary of this region toward higher values of $\Delta$. This shift reflects the fact that larger values of $A_{12}$ effectively lower the competitor density required to activate the escape response of species 1. Consequently, larger values of $\Delta$ are needed to suppress this response and enter the regime in which species 1 moves predominantly via linear diffusion.

Overall, varying the value of $A_{12}$ does not substantially alter the qualitative picture presented in the main text, apart from a modest reduction in the diversity of spatial patterns that emerge.

\subsubsection{Varying Motility Rates $\mu_0$ and $\mu_S$}

We now examine how the phase diagram is modified when varying the motility parameters, namely the linear diffusion rate $\mu_0$ and the escape-response rate $\mu_S$. In the main text, we fixed these parameters to $\mu_S=0.1$ and $\mu_0=1\times10^{-3}$, or, for analytical convenience, considered the limiting case $\mu_0\to0$.
 \begin{figure}
    \centering
    \includegraphics[width=1\columnwidth]{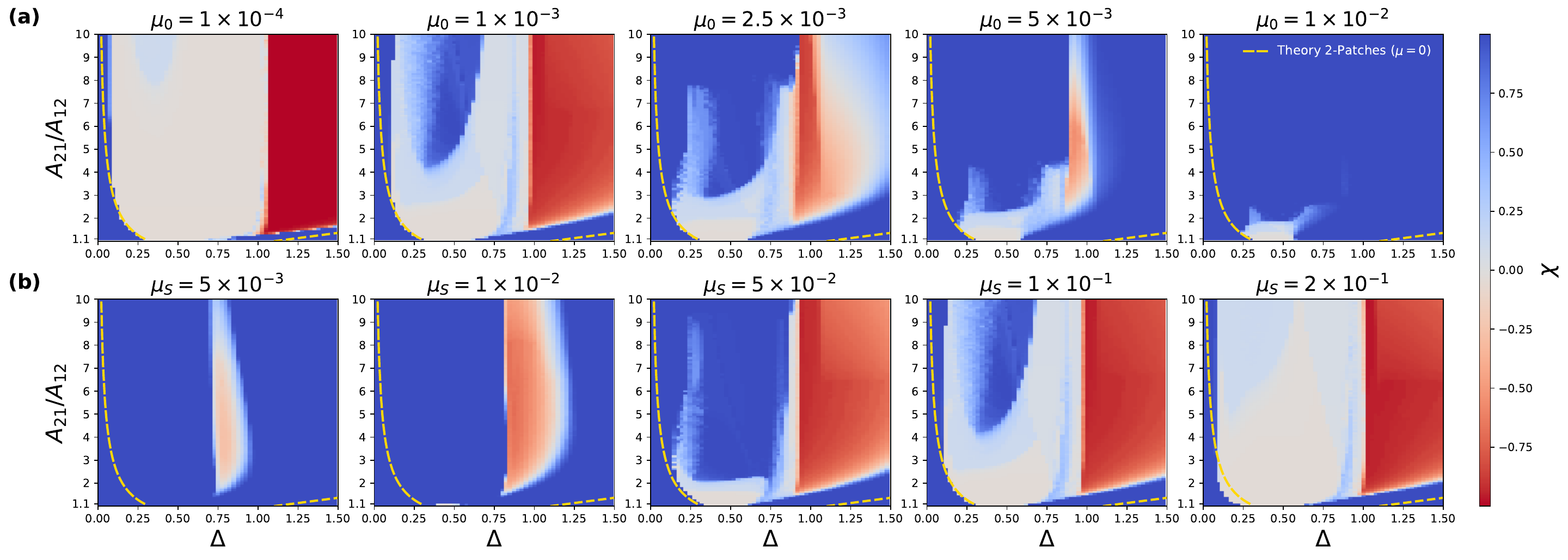}
    \caption{(a) Heatmaps showing the normalized difference between the total equilibrium abundances of species~1 and species~2, $\chi\equiv(x_1-x_2)/L^2$, where $x_i=\sum_\alpha x_i^\alpha$. Blue colors indicate dominance of species~1, red colors indicate dominance of species~2, and white or grey regions correspond to comparable abundances. The activation threshold $\Delta$ is shown on the horizontal axis and the competition ratio $A_{21}/A_{12}$ on the vertical axis. Results are shown for increasing values of the linear diffusion rate $\mu_0=10^{-4},\,10^{-3},\,2.5\times10^{-3},\,5\times10^{-3},\,10^{-2}$, while keeping the escape rate fixed at $\mu_S=0.1$. (b) Same as panel (a), but for increasing values of the escape rate $\mu_S=5\times10^{-3},\,10^{-2},\,5\times10^{-2},\,10^{-1},\,2\times10^{-1}$, with the linear diffusion rate fixed at $\mu_0=10^{-3}$. All results are obtained from numerical integration of the model with parameters $A_{11}=A_{22}=K=r=1$, $A_{12}=1.1$, $\xi=100$, and lattice side length $L=51$. Initial conditions correspond to $x_1^\alpha(0)=x_2^\alpha(0)=0.001$ in the central cell and zero elsewhere. The mobility matrices ${\bf M}$ are constructed using a Moore neighborhood.}
    \label{figS4}
\end{figure}
In panel (a) of Fig. \ref{figS4}, we examine the effect of increasing the linear diffusion rate by considering five different values of $\mu_0$, while keeping the escape-response rate fixed at $\mu_S=0.1$, as in the main text. Panel (b) instead shows the results obtained by varying $\mu_S$ over five increasing values, while fixing $\mu_0=1\times10^{-3}$. As expected, increasing linear diffusion and/or decreasing the nonlinear escape rate progressively reduces the region of parameter space in which coexistence is observed. Interestingly, the way coexistence is lost differs between the two cases. As $\mu_0$ increases, coexistence persists only for small values of $A_{21}/A_{12}$ and intermediate values of $\Delta$, where residual concentric-ring–like structures can still form. By contrast, when $\mu_S$ is decreased, coexistence survives primarily in the regime where species 2 dominates, corresponding to larger values of $\Delta$ and a broader range of $A_{21}/A_{12}$. This highlights that $\mu_0$ and $\mu_S$ affect coexistence in qualitatively different and asymmetric ways.

Beyond confirming that coexistence requires predominantly nonlinear, density-dependent motility relative to standard diffusion ($\mu_S \gg \mu_0$) in order to exploit a competition–colonization trade-off~\cite{Amarasekare}, these supplementary analyses demonstrate that coexistence can occur in multiple forms across a range of $\mu_0$ and $\mu_S$ values, including smaller values of $\mu_S$ and larger values of $\mu_0$. Therefore, although the parameter values adopted in the main text were selected to highlight the richest set of behaviors, coexistence is not restricted to a finely tuned parameter regime.

\end{document}